# An Empirical study on Mutual fund factor-risk-shifting and its intensity on Indian Equity Mutual funds.

Dr. Rajesh ADJ Jeyaprakash[1], Dr. Senthil Arasu Balasubramanian[2], Maddikera Vijay[3],
[1,2]Department of Management Studies, National Institute of Technology, Tiruchirappalli
[3]School of Minerals, Metallurgical and Materials Engineering, Indian Institute of Technology Bhubaneswar

## Abstract

Investment style groups investment approaches to predict portfolio return variations. This study examines the relationship between investment style, style consistency, and risk-adjusted returns of Indian equity mutual funds. The methodology involves estimating size and style beta coefficients, identifying breakpoints, analysing investment styles, and assessing risk-shifting intensity. Funds transition across styles over time, reflecting rotation, drift, or strengthening trends. Many Mid Blend funds remain in the same category, while others shift to Large Blend or Mid Value, indicating value-oriented strategies or large-cap exposure. Some funds adopt high-return styles like Small Value and Small Blend, aiming for alpha through small-cap equities. Performance changes following risk structure shifts are analyzed by comparing pre- and post-shift metrics, showing that style adjustments can enhance returns based on market conditions. This study contributes to mutual fund evaluation literature by highlighting the impact of style transitions on returns.

## 1. Introduction

Investment style can be defined as a natural assembly of investment disciplines that has some predictive power in explaining future variability in returns across portfolios (Christopherson et al., 1998; Maginn et al., 2007). The investment style of a mutual fund is the most significant component of the fund's business (Brown & Goetzmann, 1997; Cooper et al., 2005). Simply, style investing embarks on investment across board asset classes instead of concentrating on individual securities.(Chan et al., 2002). The investment style is the primary investment approach that promotes the fund to investors seeking a particular type of market exposure. Economic effect of Investment Style drifts and its impact was studied by (Daniel et al., 1997).

Technically, style drift can be viewed as a change in portfolio characteristics over time or as a shift in factor loadings on style risk factors. The intensity of style risk shifting plays a major role in understanding style drift in fund management. In financial literature, studies have shown a relationship between style drift, mutual fund performance and risk management and governance also (Zhang et al., 2024; Mateus et al., 2023; Misra & Mohapatra, 2017; Herrmann et al., 2016; Wahal & Yavuz, 2013). However, whether style drift implies better, or worse performance remains a long-debated topic among researchers and practitioners in portfolio management. (Brown & Goetzmann, 1997; Wermers, 2000; Andreu et al., 2018, 2019) have concerted the style drift implies absolute performance, high alpha generator and better performance respectively. This debate centers around investing styles of funds in the market. Sharma et al., (2021) recently emphasized that size and volume anomalies are fading in the long term in the Indian stock market. In the Indian Context, Low style consistency was reported in 2017 in Indian Large Cap Fund (36.65%) and Indian Mid/Small Cap Fund (49.25%) over the horizon of five years.[1] There is no proper report that examines into the elaborated the details about what type of intensity each equity mutual fund and on the risk shifting behaviour of fund managers in Indian context. This paper sheds light on the implications and trends associated with style drift within the Indian equity mutual fund. The insights contribute valuable perspectives on the dynamic nature of active fund management and its impact on investor portfolios and theoretically, the paper also contributes towards the agency theory, where the investor portfolio and fund managers may lead to conflicts (Starks, 1987).

Bhargava et al., (2001) empirically proved among 114 mutual funds across nations that equity style allocation enhanced performance of international funds managers. There are two main implications of understanding the style of the mutual fund are, style provides performance evaluation of a fund manager's stock selection skill and enhance control over the portfolio risk (Chan et al., 2002) and identified size and book-to-market are descriptors of funds style. Sometime drift from happens in a stock automatically, for instance a mid-cap stock may eventually change to large cap stock if the values of assets increase over time (Huang et al., 2011) Hence this paper extent the examination of style exposure and its intensity for long only mutual funds in the Indian Equity mutual funds.

Comer et al., (2008) has examined 462 hybrid funds in the with tenure of 2001 to 2005, verified the importance of exposure on factor style while investing in the capital market. (Ainsworth et al., 2008) studied the direction of the style drift, on 37 Australian equity mutual funds, concluded that fund managers remain committed to their stated objectives in relationship to the fund drift. Meier & Rombouts, (2009) has conducted a comprehensive study with 3799 funds for the tenure of six years and proved that to understand the intensity of the style, style rotation occupies the major factor in selection of funds managers in the equity mutual funds. When fund drift arises, style volatility is the major components to study the fund was established and concluded that low style volatility will help managers from poor performance in the fund market (Brown et al., 2011). (Wermers, 2012) a seminal paper analyzed 2892 funds and 2670 lead managers for the tenure of 1985 to 2000, comprehensively identified chasing style will provide higher fund performance. Herrmann & Scholz, (2013) has supported style-shifting performance into active and passive components and concluded that passive style shifting leads to higher fund performance and active style linked with creating portfolio based on the specific style in the market like size, profitability, value, growth and momentum etc., whereas passive style alternatively linked with specific market and stock specific characteristics. (Herrmann et al., 2016).

[1] https://www.indexologyblog.com/2018/10/23/low-style-consistency-in-large-cap-and-mid-small-cap-fund-categories/

This raises questions among investors and poses challenges for fund managers in understanding how this fading anomaly affects mutual fund performance in India from a long-term perspective, particularly concerning the intensity of factor risk shifting. This study bridges this gap by investigating the relationship between investment style, style consistency, and risk-adjusted returns of Indian equity mutual funds, contributing to the literature on their assessment.

### 1.1. The study aims to address the research questions through research objectives (RO's)

RO1 To classify funds based on the style rotation, Style Drift and Style strengthening and weakening

RO2 To understand the relationship between Intensity of risk-shifting and performance of the mutual funds.

The literature on measuring and evaluating style drift in terms of volatility consists of two main approaches: returns-based analysis (RBA) (Blake et al., 1993; Holmes and Faff, 2008; Brown et al., 2009) and holdings-based analysis (HBA), which was pioneered by Daniel et al. (1997). A key advantage of RBA is that it utilizes readily available data, but this method has notable limitations that may lead to inaccurate results. For instance, RBA's constrained regression model has a limited capacity to capture investment changes driven by style drift (Bollen and Whaley, 2009). Additionally, when a fund's name and investment strategy are not clearly articulated, RBA can become unstable and highly unreliable (Buetow et al., 2000). Another drawback is that many passive style indexes used as benchmarks in this approach lack a distinct or "pure" style definition, as they often contain overlapping stocks. However, our approach differs from other related studies such as Wermers (2012) and Cao et al. (2017). Unlike Wermers (2012), who separately analyses "active drift" and "passive drift," our study focuses on active drift while controlling for changes in stock characteristics related to size, book-to-market ratio, and momentum. Cao et al. (2017) measure style drift by classifying a fund's investment style based on stock membership in the Russell 1000 or Russell 2000 indexes. However, we do not adopt this method, as there are no well-defined or widely accepted Chinese indexes that accurately align with fund investment styles. The key feature of the study are first, using a unique dataset that is free from survivorship bias, we provide strong evidence of intentional style drift among the sample equity funds. Our findings reveal that the majority of fund managers select stocks that deviate from the size and value/growth characteristics outlined in their investment prospectuses. This confirms the existence most both in its basic form and in terms of volatility over time.
Second, our results show that funds with higher net inflows tend to exhibit the most significant style drift during the study period.

Additionally, we find that managers of large funds with substantial assets under management (AUM) are more inclined to engage in style drift. Third, our findings indicate that style drift negatively impacts a manager's ability to select superior stocks, reducing fund returns by approximately 1.32% to 2.58%. Consistent with Brown et al. (2015)'s findings on the U.S. fund market, our study shows that dedicated funds, in contrast, tend to match their benchmark returns and exhibit lower portfolio risk. Contributions to Literature includes, it contributes to the existing literature by providing new insights into the dynamics and consequences of undetected risk-taking associated with style drift in a market with in-house fund managers. Our study enhances the understanding of how AUM-linked compensation influences fund managers' motivation and performance—an area with limited prior research (Chen et al., 2013). Additionally, the fund classification and style drift metrics we develop for the Chinese market introduce novel methodological approaches and standards. These contributions help improve transparency and clarity regarding product attributes, benefiting market participants and regulators by reducing agency conflicts between fund investors, fund managers, and other key stakeholders

## 2. Data and Methodology

### 2.1. Data

The study data set consists of 34 equity funds which exist in the Indian mutual fund market since 2006. The sample period of the study spans from January 2006 to June 2023. Based on the daily NAV value we calculated the return of the inclusive of dividend and other information. To establish a style category for the fund in the sample in the month t and place it in the relevant style box. The selected funds the study is in table 1. We have adopted (Carhart, 1997)which is most widely used in the style investing literature.

### 2.2. Tools adopted for the study:

Table I. List of Mutual Schemes and its benchmark

| Name of Scheme | ID | Type of Fund | Benchmark |
|---|---|---|---|
| Aditya Birla SL Equity Advantage Fund | Fund 1 | Large and Midcap | S&P BSE 250 Large MidCap TRI |
| Aditya Birla SL Flexi Cap Fund | Fund2 | Large Cap Fund | NIFTY 500 TRI |
| Aditya Birla SL Midcap Fund(G) | Fund 3 | Large Cap Fund | NIFTY 100 TRI |
| Aditya Birla SL Frontline Equity Fund | Fund 4 | Large Cap Fund | Nifty 100 TRI |
| Baroda BNP Paribas Multi Cap Fund | Fund 5 | Midcap | Nifty Midcap 150 TRI |
| Baroda BNP Paribas Mid Cap Fund-Reg(G) | Fund 6 | Multi Cap Fund | Nifty 500 Multicap 50:25:25 TRI |
| Baroda BNP Paribas Multi Cap Fund-Reg(G) | Fund 7 | Large & Mid-Cap | NIFTY Large Midcap 250 TRI |
| Canara Rob Emerg Equities Fund-Reg(G) | Fund 8 | Flexi Cap Fund | S&P BSE 500 TRI |
| Canara Rob Flexi Cap Fund-Reg(G) | Fund 9 | Large & Mid-cap | NIFTY Large Midcap 250 (TRI |
| DSP Equity Opportunities Fund-Reg(G) | Fund 10 | Flexi Cap fund | NIFTY 500 TRI |
| DSP Flexi Cap Fund-Reg(IDCW) | Fund 11 | Large Cap Fund | Nifty 100 |
| Franklin India Bluechip Fund(G) | Fund 12 | Large and Midcap | Nifty Large Midcap 250 |
| Franklin India Equity Advantage Fund(G) | Fund 13 | Flexi Cap Fund | Nifty 500 |
| Franklin India Flexi Cap Fund(G) | Fund 14 | Midcap | Nifty Midcap 150 |
| Franklin India Prima Fund(G) | Fund 15 | Flexi Cap Fund | NIFTY 500 Total Returns Index |
| HDFC Flexi Cap Fund(G) | Fund 16 | Large Cap Fund | NIFTY 100 |
| HDFC Top 100 Fund(G) | Fund 17 | Flexi Cap Fund | NIFTY 500 TRI |
| HSBC Flexi Cap Fund(G) | Fund 18 | Large Cap Fund | Nifty 100 TRI |
| HSBC Large Cap Equity Fund(G) | Fund 19 | Large and Midcap | Nifty Large MidCap TRI |
| ICICI Pru Midcap Fund(G) | Fund 20 | Midcap | NIFTY Midcap 150 TRI |
| ICICI Pru Large & Mid Cap Fund(G) | Fund 21 | Large Cap Fund | S&P BSE 100 TRI |
| JM Large Cap Fund-Reg(G) | Fund 22 | Large Cap Fund | Nifty 100 TRI |
| Kotak Bluechip Fund(IDCW) | Fund 23 | Large and Midcap | Nifty Large Midcap 250 TRI |
| Kotak Equity Opp Fund(G) | Fund 24 | Large and Midcap | Nifty Large Midcap 250 TRI |
| L&T Large and Midcap Fund-Reg(G) | Fund 25 | Midcap | NIFTY Midcap 150 TRI |
| L&T Midcap Fund-Reg(G) | Fund 26 | Midcap | NIFTY Midcap 150 TRI |
| LIC MF Large Cap Fund-Reg(G) | Fund 27 | Large Cap Fund | Nifty 100 TRI |
| Nippon India Growth Fund(G) | Fund 28 | Midcap | NIFTY Midcap 150 TRI |
| Nippon India Multi Cap Fund(G) | Fund 29 | Multi Cap Fund | Nifty 500 Multicap 50:25:25 Total Return Index |
| Nippon India Vision Fund(G) | Fund 30 | Large and Midcap | S&P BSE 250 Large MidCap Total Return Index |
| PGIM India Large Cap Fund(G) | Fund 31 | Large Cap Fund | NIFTY 100 Total Return Index |
| SBI BlueChip Fund-Reg(G) | Fund 32 | Large Cap Fund | S&P BSE 100 Index |
| SBI Magnum Midcap Fund-Reg(G) | Fund 33 | Midcap | NIFTY Midcap 150 Total Return Index |
| Sundaram Multi Cap Fund(G) | Fund 34 | Multi Cap Fund | Nifty 500 Multicap 50:25:25 Total Return Index |

### 2.2. Methodology

The process of evaluating the relationship between risk-shifting of mutual funds and their risk-adjusted performance involves several interconnected stages.

The initial phase involves the identification of mutual funds exhibiting breaks in style risk. It includes the calculation of Beta Coefficients for each factor loading within the fund, specifically $\beta 1$, $\beta 2$, and $\beta 3$. The consolidation of $\beta 2$ and $\beta 3$ into a cohesive file is essential for the subsequent steps. Furthermore, this stage employs the Bai and Perron Model to identify structural breaks in time series data, specifically focusing on style risk. The goal is to calculate the discrepancy between a fund's style risk and the style risk of its benchmark index. The consideration here is exclusively given to style risk breaks that surpass those inherently embedded within the funds' benchmark indices.

The subsequent phase involves categorizing the mutual funds into three distinct groups based on the severity of their style risk changes: Extreme: Indicates a significant style rotation. Moderate: Denotes a moderate degree of style drifting. Weak: Refers to either a style-strengthening or weakening trend.

In the latter phase, the risk-adjusted performance of each fund is determined. This is achieved by adjusting the fund's returns concerning its risk exposure using established risk-adjusted measures such as the Sharpe ratio or Treynor ratio.

The final phase entails analysing the association between risk-shifting and risk adjusted performance. It compares the risk-adjusted performance of funds demonstrating style risk breaks against those that do not exhibit such breaks. The objective here is to identify factors contributing to the relationship between risk-shifting behaviours and the resulting impact on risk-adjusted performance.

This comprehensive methodology aims to systematically assess how changes in mutual funds' style risks influence their risk-adjusted performance, providing insights into the relationship between style drifts and fund performance, thereby aiding in investment decisions.

**Steps involved in the study:**

1. Estimation of size and style beta coefficient in the sample funds in the study.
2. Breakpoints in funds' benchmark adjusted size and style betas.
3. The investment styles of funds.
4. The intensity of risk-shifting.

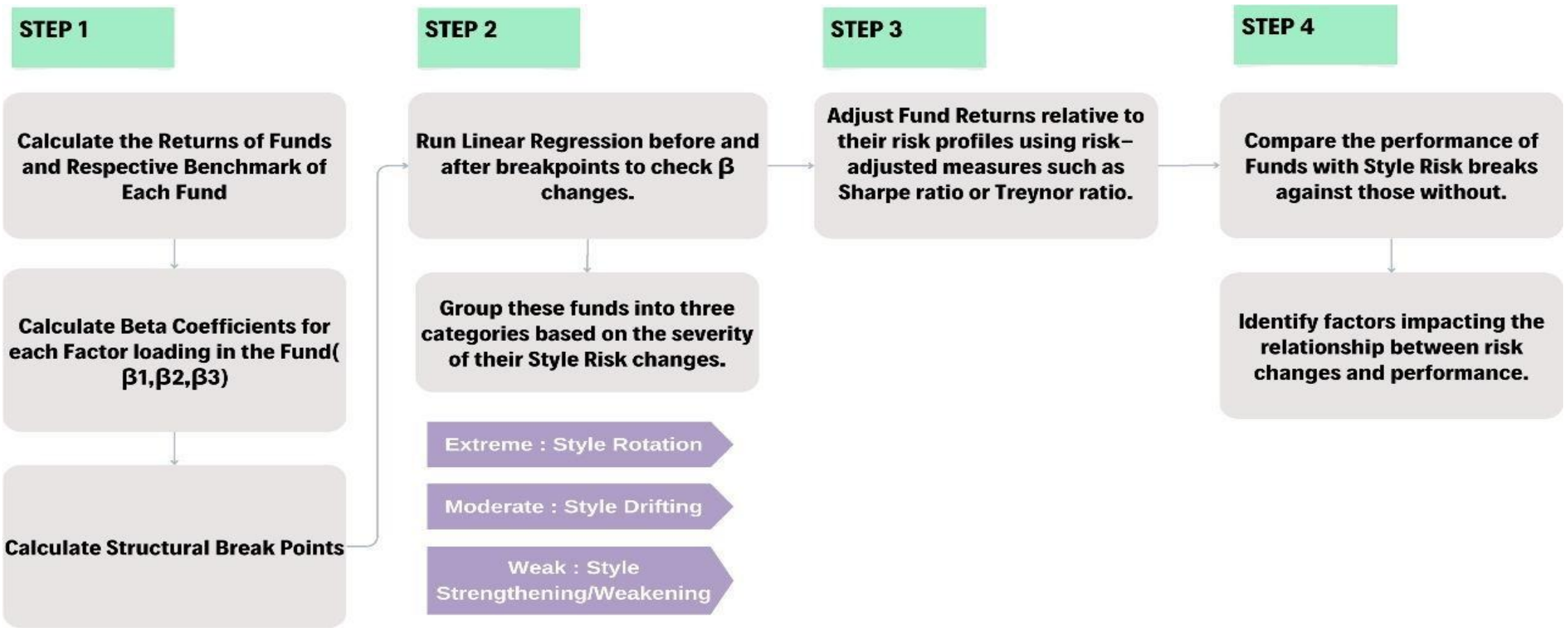

The process of evaluating the relationship between risk-shifting of mutual funds and their risk-adjusted performance involves several interconnected stages. The initial phase involves the identification of mutual funds exhibiting breaks in style risk. It includes the calculation of Beta Coefficients for each factor loading within the fund, specifically β1, β2, and β3. The consolidation of β2 and β3 into a cohesive file is essential for the subsequent steps. Furthermore, this stage employs the Bai and Perron Model to identify structural breaks in time series data, specifically focusing on style risk. The goal is to calculate the discrepancy between a fund's style risk and the style risk of its benchmark index. The consideration here is exclusively given to style risk breaks that surpass those inherently embedded within the funds' benchmark indices. The subsequent phase involves categorizing the mutual funds into three distinct groups based on the severity of their style risk changes: Extreme: Indicates a significant style rotation. Moderate: Denotes a moderate degree of style drifting. Weak: Refers to either a style-strengthening or weakening trend.

### 2.3 Investment style

This study examines the style drift by understanding the volatility in the fund's portfolio based on the style adoption namely size, growth/value, and momentum. This study adopted recent previous studies (Wermers, 2011; K. C. Brown et al., 2015; Mateus et al., 2023). Our focus on the all the investment style based on the theory (Carhart, 1997). Our study is differentiated from (Cao et al., 2017) where they adopted only size as the aspect in investment style.

### 2.4 Intensity of Risk-shifting

Classifying fund based on the intensity of their benchmark-adjusted factor-risk-shifting into three categories (Mateus et al., 2023)

| Nature of Intensity | Description of the Intensity | Remarks |
|---|---|---|
| Style Rotation funds | Beta coefficient of SMB or HML remains significant before (time t) and after the break (t+1) but change the sign. | Classified funds are rated as the highest level of risk-shifting. |
| Style Drifting funds | Beta Coefficient of SMB or HML goes from significant (Positive or negative) to insignificant – or vice versa – before and after regime. | Natured as moderate risk-shifting category. |
| Style strengthening/weakening funds | Only factor loading increases or decrease before and after the break but remain the same sign and significance. | Natured as least risk-shifting. |

## 3. Empirical results:

We begin our empirical investigation with the examination of the number of breaks in investment style extremity to understand the total breaks in the fund and the quantify the extent to which follows specific styles.

### 3.1. Understanding Break Points in the Funds

**Table No. 1 Breakpoints in funds benchmark-adjusted size and style betas**

| Number of Breaks | Number of Funds | Total Breaks |
|---|---|---|
| 1 | 34 | 34 |
| 2 | 31 | 62 |
| 3 | 32 | 96 |
| 4 | 34 | 136 |
| 5 | 29 | 145 |
| Total | 160 | 473 |

Table no 1 shows the breakdown of funds per number of structural breaks. Table 1 outlines the breakdown of funds per number of structural breaks. It shows that 160 funds in our sample have at least one structural change in risk and the total number of regime changes in our sample is 473. The vast majority (29) of those funds have five structural breaks.

In this study, by identifying structural changes in the benchmark adjusted alpha, we have eliminated a considerable number of what we consider false changes in size/style, risk exposure, driven by changes in stock characteristics reflected in the market index used as a benchmark (S&P500). Specifically, should we have used the standard Fama-French three-factor model to identify the breakpoints instead of the AGT, there would have been 160 funds with at least one structural break(s) in size/style betas, a result much closer to the Annaert and Van Campenhout (2007) study that uses daily data and finds at least one break in all the funds in their sample.

### 3.2. The investment styles of funds and Intensity of Risk-Shifting

Using the sign and significance of the SMB and HML factor loadings, in each risk regime, we classify the funds in our sample into one of the nine style categories from the Morningstar Style Box, obtained as combinations of three style categories (value, blend, and growth) and three size categories (small-, medium- and large-cap). The purpose of identifying fund’s style exposure before and after the structural break in their benchmark-adjusted beta occurs is to identify the funds that have significantly shifted their risk before and after each break has occurred. Hence, if after identifying a breakpoint, the SMB coefficient is showing that the mutual fund’s SMB beta has progressed from significantly positive before the break to significantly negative after it, for example, it implies this has fund changed its size exposure from small-cap to large-cap stocks significantly altering its risk profile. Similar can be said for the changes in coefficients associated with HML factor. The shifts in SMB and HML beta coefficients may result in a complete rotation in style.

In this paper, we define the level of intensity of risk shift not by looking at the absolute change in factor loadings (as in Herrmann, Rohleder, and Scholz 2016) before and after the break but we account for the sign and the significance of the loadings. Consider, for instance, a fund that increases the beta by 0.3; if the beta remains of the same sign and significance, the fund is placed in the same Morningstar style (risk) category, so the investor would not view such factor risk change as extreme. One the other hand, a change in betas of 0.3 that is accompanied by the change in sign and significance (say SMB beta changes from +0.1 to −0.8, both significant), implies that a manager changes exposure from small to large cap stocks, which for a fund branded as a ‘small cap fund’ is a cause for concern among investors. Hence, once the structural break(s) for each fund has been established, we determine the funds investment style by estimating the FF3 model in the regime prior to (period t) and post (period t+1) structural change. The level of change in the sign and significance of the estimated SMB and HML factor loadings between the regimes infers a different degree of shifting of the fund’s size and style risk. To which extent the fund changes its risk exposure following the benchmark-adjusted break depends on the type of change in SMB and HML beta coefficients compared to the regime before the break.

To this end, we group the funds according to the intensity of their benchmark-adjusted factor-risk-shifting into three categories, namely: 1) Style rotation funds: those whose SMB or HML beta(s) remain significant before (time t) and after the break (time t+1), but change the sign, indicating that the fund has changed the style from one regime to another. Note that funds that exhibit more than one structural break can change style more than once over the sample period. The success of such a strategy where fund managers rotate between the styles has been well documented in the financial literature. Evidence on successful style rotation strategies in the US market can be found in Kao and Schumaker (1999) and Asness et al. (2000) among others and is not contained to the US market only.12 However, the evidence also highlights that a typical mutual fund would have risk constraints that would prevent them from exploiting the full benefits of style rotation. Therefore, in our sample, we consider that funds experiencing style rotation alter their risk characteristics the most, and consequently, we classify them as funds with the highest level of risk-shifting. 2) Style drifting funds: those whose beta goes from significant (positive or negative) to insignificant – or vice versa – before and after the regime change. We place those funds in the moderate risk-shifting category and Table 2.

The number of breakpoints in factor loadings, per style and intensity of style change and acknowledge that this type of change can be a result of a natural drift in a fund (e.g. stocks in a small-cap fund gradually getting larger and as a result the fund appears to be drifting towards mid-cap category, while no actual change in holdings has occurred).However, note that those natural changes will be affecting factor loadings of the fund's benchmark as well, so by applying benchmark-adjusted model to identify structural changes in style betas, we isolate those changes that are the result of true drift by fund managers beyond that of the benchmark.
3) Style strengthening/weakening funds: those whose factor loadings increase or decrease before and after the break but remain of the same sign and significance. This can happen because of portfolio manager's picking of more (or less) extreme stocks within the same style or naturally (e.g. if the average P/E and market value of the portfolio increases/decreases over time). Similarly, as in (2) the funds exhibiting a natural shift that is also embedded in the benchmark will be eliminated and only those whose fund managers make active increases/decreases in portfolio size or P/E ratio, for instance, will be captured. We consider this to be the least intense risk-shifting, as the fund remains within the same style classification in both regimes and would agree with investor's initial risk preferences.

Table 2 shows the number of risk changes that result in style rotation (darkest shade), style drifting (lightest shade) and style strengthening/weakening (medium shade, diagonal of the matrix) from period t to t+1, i.e. before and after each structural break.

The table is organised as a matrix, where rows represent style of fund in period t (before breakpoint) and columns are style of fund in (after breakpoint) and should be interpreted as follows. For instance, the first cell in the table shows that in 19 instances in our sample period funds that were classified as Large Value funds before the break – have strengthened or weakened their style following the break in benchmark-adjusted factor loading, remaining in the Large Value category. Then, the cell below shows that in 15 instances funds that were classified as Large Blend in period t changed their style to Large Value in period t+1 following the break. And so on. It becomes evident that very few changes in style risk exposure (9 in total) result in style rotation. That is good news for investors, as style switching implies a significant shift in risk of the fund, often outside the risk parameters, the fund's official investment style corresponds to, 927 out of 1558 style changes result in what we classify in Section 2.2.3 as style drift, while 622 changes are reflecting strengthening or weakening existing style exposure (obtained as the sum of the values in the highlighted matrix diagonal in Table 2). The table also reveals some less desirable news for investors: it shows that the fund managers seem not to apply what the literature on style investing suggests. Specifically, the vast majority style changes in Table 2 show that funds move towards the mid-cap or blend style, rather than into pure value or growth, small-cap or large-cap styles, which are proven to perform very well historically in different periods as mentioned in Section 1 of this paper. Note that Table 2 includes all benchmark-adjusted risk shifts in our sample, even those estimated using betas based on short break periods, as explained in the previous section. Where relevant in the analysis, we will test the validity of our results by removing risk shifts based on breaks shorter than 24 months in an attempt to remove bias stemming from potential error in beta estimation over short periods.

### 3.3. Intensity of Risk-Shift among Various styles of Mutual funds:

**Table No. 2 Intensity of Risk-Shift and Categories**

| t\t+1 | Large Value | Large Blend | Large Growth | Mid Value | Mid Blend | Mid Growth | Small Value | Small Blend | Small Growth | Total |
|---|---|---|---|---|---|---|---|---|---|---|
| Large Value | 15 | 19 | 0 | 18 | 14 | 0 | 1 | 0 | 0 | 67 |
| Large Blend | 22 | 72 | 11 | 16 | 70 | 5 | 0 | 4 | 0 | 200 |
| Large Growth | 2 | 9 | 5 | 2 | 13 | 2 | 0 | 1 | 0 | 34 |
| Mid Value | 10 | 17 | 0 | 44 | 50 | 1 | 16 | 10 | 1 | 149 |
| Mid Blend | 10 | 53 | 16 | 59 | 333 | 61 | 21 | 52 | 16 | 621 |
| Mid-Growth | 1 | 6 | 8 | 6 | 67 | 38 | 23 | 56 | 17 | 171 |
| Small Value | 0 | 2 | 0 | 8 | 45 | 18 | 24 | 56 | 11 | 67 |
| Small Blend | 0 | 0 | 1 | 2 | 10 | 21 | 2 | 17 | 32 | 164 |
| Small Growth | 0 | 1 | 1 | 7 | 14 | 0 | 23 | 19 | 2 | 85 |
| Total | 60 | 179 | 42 | 162 | 616 | 146 | 110 | 215 | 78 | 1558 |

**Table 3. No breaks funds vs. funds with structural breaks: performance indicators.**

| Funds with… | Number of funds (breaks) | Excess return p.a.(%) | Std.Dev. p.a. (%) | Sharpe ratio p.a. | FF3 alpha p.a. (%) | AGT alpha p.a. (%) |
|---|---|---|---|---|---|---|
| No breaks | 41 | 3.45 | 18.74 | 0.26 | -0.90 | -0.98 |
| 1 break | 34(34) | 5.53 | 17.49 | 0.34 | -0.19 | -0.27 |
| 2 breaks | 31(62) | 7.89 | 17.67 | 0.59 | 0.60 | 0.49 |
| 3 breaks | 32(96) | 7.64 | 16.67 | 0.49 | 0.59 | 0.19 |
| 4breaks | 34(136) | 5.70 | 15.23 | 0.4 | 0.45 | -0.78 |
| 5 breaks | 29(145) | 5.88 | 18.42 | 0.28 | 0.28 | -0.39 |
| All funds with breaks | 160(473) | 5.89 | 18.16 | 0.38 | 0.30 | -0.35 |

First, we analyse the transition of funds across investment styles from period *t* to period *t+1* as shown in Table 2. The matrix highlights how mutual funds shift their style classifications over time—whether they maintain their original style or move to a different category, thereby indicating possible style rotation, style drift, or style strengthening/weakening. For example, a significant number of funds originally classified as *Mid Blend* at time *t* remained in the same category at *t+1* (333 funds), but many also shifted to adjacent styles such as *Large Blend* (53 funds) and *Mid Value* (59 funds), suggesting style drift towards value-oriented strategies or larger capitalization exposure.

Second, we evaluate whether specific patterns of risk-shifting—such as movements into styles associated with historically higher returns like *Small Value*—contribute to performance improvements following a change in risk structure. Notably, a moderate inflow of funds into *Small Value* and *Small Blend* categories can be observed, indicating that some funds may be seeking alpha through increased exposure to small-cap equities, which may reflect style rotation into riskier segments for potential excess returns.

Third, we assess whether the performance of funds exhibiting risk-shifting behavior is linked to specific style transitions. For instance, funds that moved into *Small Value* or *Mid Growth* might experience varying levels of return enhancement, depending on market conditions favoring small-cap or growth-oriented styles during *t+1*. By comparing pre- and post-shift performance metrics for each style pair in Table 2, we can identify whether such shifts are associated with material improvements in fund performance, potentially indicating successful style rotation or strengthening.

Finally, we analyze the top 10% of funds by excess returns and compare them against the bottom decile to understand if superior performance correlates with distinct risk-shifting behavior. If top-performing funds are disproportionately represented among those shifting into high-return styles such as *Small Value* or *Mid Growth*, this could imply that proactive style adjustments contribute significantly to performance outperformance. Conversely, bottom-decile funds might exhibit either minimal risk-shifting or unsuccessful attempts at style rotation, thereby reinforcing the importance of strategic risk realignment in achieving superior risk-adjusted returns.

The data presented evaluates mutual fund performance based on the number of risk structure breaks, representing significant changes in a fund's investment style or risk exposure over time. The objective is to determine whether such risk-shifting behaviour—through style rotation, drift, or strengthening/weakening—leads to improved fund performance.

Funds with no risk structure breaks (41 funds) demonstrate moderate annual excess returns of 5.78% and relatively high volatility (18.74%), resulting in a low Sharpe ratio of 0.26. More notably, these funds show negative alpha values relative to both the Fama-French 3-factor model (-0.90%) and the AGT model (-0.98%), indicating that their performance consistently lags behind benchmark expectations. This suggests that funds maintaining a static risk profile, without adapting to market dynamics, struggle to generate superior risk-adjusted returns and are likely penalized in shifting market environments due to a lack of flexibility.

Funds with one structural break (12 funds) exhibit a slightly lower return (5.53%) but with lower volatility (17.49%), improving their Sharpe ratio to 0.34, alongside less negative alphas (-0.19% FF3, -0.27% AGT). While the return is slightly lower than the no-break group, the enhanced efficiency in risk-adjusted terms and reduced underperformance suggest that a single, well-timed adjustment in investment style or risk exposure can be beneficial, especially in optimizing performance relative to risk taken.

The most compelling findings arise from funds that experience two to three structural breaks. These funds report the highest annual excess returns—7.89% for two breaks and 7.64% for three breaks—while also maintaining lower volatility (approximately 16.67%–17.67%). Consequently, they achieve significantly higher Sharpe ratios (0.59 and 0.49, respectively). Crucially, these funds deliver positive alpha values (0.60% and 0.59% FF3 alpha), suggesting consistent outperformance relative to market benchmarks. This indicates that moderate risk-shifting, where funds tactically adjust their investment styles in response to market conditions, is strongly associated with enhanced performance—both in absolute terms and on a risk-adjusted basis. These funds appear to successfully implement style rotation or strengthening, improving their market positioning and capitalizing on favourable style exposures.

In contrast, funds with four or five risk structure breaks show diminishing returns from risk-shifting. While four-break funds post average returns (5.70%) and a modest Sharpe ratio (0.40), their alpha is mixed—positive by FF3 (0.45%) but negative by AGT (-0.78%), indicating inconsistent outperformance. Funds with five breaks (286 funds) report 5.88% returns and Sharpe ratio of 0.28, only slightly better than the no-break group, with marginally positive FF3 alpha (0.28%) and negative AGT alpha (-0.39%). These results imply that excessive risk-shifting may not yield superior performance, potentially due to overreaction to market signals or lack of coherent strategy, leading to instability or higher transaction costs that erode returns.

When aggregating all funds with breaks (160 funds), their average excess return (5.89%) is slightly higher than that of static funds, but with lower volatility (18.16%) and an improved Sharpe ratio (0.38). The FF3 alpha (0.30%) is marginally positive, while AGT alpha remains slightly negative (-0.35%). These aggregate results suggest that risk-shifting, in general, leads to better risk-adjusted outcomes than maintaining a fixed style, but the effectiveness of risk-shifting depends significantly on its frequency and strategic execution.

In summary, the data clearly illustrates that moderate, well-timed risk-shifting (2–3 breaks) is positively correlated with superior fund performance, both in absolute returns and relative to risk. However, excessive risk structure changes (4–5 breaks) may offer diminishing or negligible benefits, highlighting the importance of strategic and judicious adjustments rather than frequent or erratic changes. Funds that can adapt effectively to market conditions through targeted style shifts appear best positioned to outperform, while static funds lag due to their inability to respond to evolving market opportunities.

4. Findings:

Funds exhibit various transitions across investment styles over time, as shown in Table 2, indicating the presence of style rotation, style drift, or style strengthening/weakening. A significant number of funds originally classified as Mid Blend remain in the same category, while others shift to styles like Large Blend or Mid Value, suggesting a drift toward value-oriented strategies or larger capitalization exposure. Some funds actively move into historically high-return styles, such as Small Value and Small Blend, possibly seeking alpha through increased exposure to small-cap equities. Performance improvements following risk structure changes are analyzed by comparing pre- and post-shift metrics, revealing that style shifts can sometimes enhance returns depending on market conditions. Top-performing funds are often those that shift into high-return styles, such as Small Value or Mid Growth, suggesting that proactive style adjustments contribute to superior performance. Conversely, bottom-decile funds either exhibit minimal risk-shifting or fail in their attempts at style rotation, emphasizing the importance of strategic risk realignment. Funds with no risk structure breaks (914 funds) generate moderate excess returns (5.78%) but have high volatility (18.74%) and low Sharpe ratios (0.26), indicating weak risk-adjusted performance. These static funds also show negative alpha values (-0.90% FF3, -0.98% AGT), implying underperformance relative to market benchmarks. Funds with one risk structure break (12 funds) have slightly lower returns (5.53%) but improved risk efficiency, with a higher Sharpe ratio (0.34) and less negative alpha values, indicating a beneficial impact of a single, well-timed adjustment. Funds experiencing two to three structural breaks achieve the highest excess returns (7.89% for two breaks, 7.64% for three breaks) with lower volatility (16.67%–17.67%) and significantly higher Sharpe ratios (0.59 and 0.49). These funds also generate positive alpha values (0.60% and 0.59% FF3), demonstrating strong outperformance and suggesting that moderate, tactical risk-shifting enhances fund performance. Funds with four or five risk structure breaks display diminishing returns, with mixed alpha values and inconsistent performance improvements, implying that excessive risk-shifting may lead to instability or increased transaction costs. Aggregated results for all funds with breaks (367 funds) show slightly higher excess returns (5.89%) and improved risk-adjusted outcomes, though the effectiveness of risk-shifting depends on its frequency and execution.

5. Conclusion:

Moderate, well-timed risk-shifting (2–3 breaks) is positively correlated with superior fund performance in both absolute returns and risk-adjusted terms. Excessive risk structure changes (4–5 breaks) tend to offer diminishing or negligible benefits, possibly due to overreaction to market signals or increased transaction costs. Funds that strategically adjust their investment styles in response to market conditions are better positioned to outperform. In contrast, static funds that do not adapt to market changes tend to underperform due to their lack of flexibility. Overall, the study highlights the importance of a balanced approach to risk-shifting, where strategic and judicious adjustments lead to better performance, while frequent or erratic changes may undermine returns.

6. Practical Implications:

From the analysis and result discussion it is very clear that a fund managers should avoid working with static investment style, it is recommended that a manager can adopt moderately 2 to 3 style transitions in the balancing activity which in turn will reduce underperformance in the market. Fund managers has to implement any style exposure with market condition in mind. Excessive risk-shifting can lead to uneven returns and it will also increased the transaction costs. Among the funds studied, it is revealed that style rotations through strong fundamental analysis will read to overreacting to short term market fluctuations. To reduce risk-shifting a investor can exchange risk management capabilities in the AMC's. Managers should focus on styles that align with historical risk-return profiles, such as selectively increasing exposure to small-cap or growth-oriented equities when market conditions are favourable.